\documentclass[prl,twocolumn,showpacs,preprintnumbers,amsmath,%
amssymb,superscriptaddress,floatfix]{revtex4}

\usepackage{graphicx}
\usepackage{dcolumn}
\usepackage{bm}
\usepackage[latin1]{inputenc}
\usepackage{color}

\newcommand{\ZZ}{\mathbb Z}

\begin{document}

\title{Symmetries shape the current in ratchets induced by a bi-harmonic force}

\author{Niurka R.\ Quintero}
\email{niurka@us.es}
\affiliation{Departamento de F\'\i sica Aplicada I, E.\ U.\ P.,
Universidad de Sevilla, Virgen de \'Africa 7, E-41011, Sevilla, Spain}

\author{Jos\'e A.\ Cuesta}
\email{cuesta@math.uc3m.es}
\affiliation{Grupo Interdisciplinar de Sistemas Complejos (GISC), Departamento
de Matem\'aticas, Universidad Carlos III de Madrid, avda de la Universidad 30,
E-28911 Legan\'es, Madrid, Spain}

\author{Renato Alvarez-Nodarse}
\email{ran@us.es}
\affiliation{Departamento de An\'alisis Matem\'atico, Universidad de Sevilla,
apdo 1160, E-41080, Sevilla, Spain}

\date{\today}

\large

\begin{abstract}
Equations describing the evolution of particles, solitons, or localized structures,
driven by a zero-average, periodic, external force, and invariant under time
reversal and a half-period time shift, exhibit a ratchet
current when the driving force breaks these symmetries. 
The bi-harmonic force $f(t)=\epsilon_1\cos(q
\omega t+\phi_1)+\epsilon_2\cos(p\omega t+\phi_2)$ does it
for almost any choice of $\phi_{1}$ and $\phi_{2}$, provided $p$ and $q$
are two co-prime integers such that $p+q$ is odd. It has been widely observed,
in experiments in semiconductors, in Josephson-junctions, photonic crystals, etc., as well as in
simulations, that the ratchet current induced by this force has the shape
$v\propto\epsilon_1^p\epsilon_2^q\cos(p \phi_{1} - q \phi_{2} + \theta_0)$ 
for small amplitudes, where $\theta_0$ depends on the damping ($\theta_0=\pi/2$
if there is no damping, and $\theta_0=0$ for overdamped systems). We
rigorously prove that this precise shape can be obtained
solely from the broken symmetries of the system and is independent of the
details of the equation describing the system.
\end{abstract}


\maketitle

Ratchet-like transport phenomena, where a net motion of particles or
solitons is induced by zero-average forces, can be observed in many
physical systems. Such is, for instance, the dc current in 
semiconductors \cite{ieee,new28,epl99}, the net motion of fluxons in long
Josephson Junctions (JJs) \cite{usti,nori},
of vortices in superconductors \cite{vicent}, of cold atoms in optical
lattices \cite{schia,gommers}, or the rectification of Brownian motion
\cite{mm,app,brownian}. In some of these systems, 
the ratchet-like motion is induced by means of spatial asymmetries 
\cite{physicstoday,rei}. In the others the 
transport can also appear if some temporal symmetries are broken by
time-dependent forces e.g. \cite{adjari,rei,asym2,rei2,sazo,cole}.
This latter case has two advantages:
It is generally easier to analyze theoretically, and it is more amenable to
experimental observation e.g.\ in semiconductors \cite{new28}, 
in JJs \cite{usti,nori} or in optical lattices
\cite{gommers,renzoni3}.  

A big deal of simulations and experiments have show
\cite{usti,nori,schia,gommers,brownian,chaos,rei3,goychuk,
new27b,morales,gorbach,marchesoni86,new28,taylor}
that in many different systems
the behavior of the ratchet velocity $v$ driven by the $T$-periodic   
bi-harmonic force
\begin{eqnarray} \label{eqf}
f(t)=\epsilon_{1} \cos(q\omega t +\phi_{1})+
\epsilon_{2} \cos(p\omega t +\phi_{2}),
\end{eqnarray}
where $T=2 \pi/\omega$, $\phi_{1}$ and $\phi_{2}$ are the phases,   
$p$ and $q$ are co-primes with $p+q$ odd, and
the amplitudes $\epsilon_1$ and $\epsilon_2$ are small, is given by
the expression
\begin{equation}
v=B\epsilon_1^p\epsilon_2^q \cos(p\phi_1- q\phi_2 +\theta_0),
\label{eq:vexp}
\end{equation}
where $B$ and $\theta_{0}$ depend on the parameters of the model and on $\omega$
but neither on the amplitudes nor on the phases 
\cite{usti,gommers,sazo,renzoni3,chaos,rei3,new27b}. It has also been
shown for specific systems  
that nondissipative dynamics have $\theta_0=\pi/2$ \cite{schia,asym2}, 
whereas overdamped ones have $\theta_0=0$ \cite{nori,gorbach,marchesoni86}. 
The aim of this
paper is to show that symmetry considerations alone are enough to predict
the behavior (\ref{eq:vexp}). This is a strong result because it is valid
for \emph{any} equation that describes the system, no matter the type of
nonlinear terms it may contain, as long as it shows invariance under certain
symmetry transformations ---which will state precisely below.

Attempts at determining the shape of the current \eqref{eq:vexp} can
be found even in the pioneering works \cite{taylor,new28}, aimed at developing a
sensitive method of measuring deviations from Ohm's law.
Their analysis, however, relies on an expansion of $v$ in odd moments of $f(t)$,
justified by the adiabatic response of the system to an applied field
(see also \cite{taylor2}).
While it cannot be ruled out that such an expansion holds for some systems,
or in this adiabatic limit, it is certainly not valid in general.
In fact, if one applies that expansion to
related dissipative systems, like those of 
Refs.~\cite{usti,schia,gommers,asym2,sazo,chaos,rei3,new27b,new28},
the value $\theta_0=0$ is always obtained, whereas $\theta_{0}\ne 0$
in general ---it can even be $\theta_0=\pi/2$ when dissipation vanishes.
We illustrate this fact by analyzing in the Supplementary Material an
exactly solvable example. There one can readily see that the moment
expansion is in general an incorrect assumption; only in the overdamped or
the adiabatic limits this expansion becomes correct, but we do not know
of any proof that this holds for systems other than this specific example.

Let $E[x(t),f(t)]=0$ denote a functional equation (which can represent
an ordinary or partial differential equation, an integral equation,
etc.) describing the evolution of a
particle, soliton, or localized structure whose position is given by $x(t)$,
under the driving of a zero-average, external, periodic
force $f(t)=f(t+T)$, $T>0$. One such system is said to have ratchet-like
behavior if the average velocity, defined as \cite{rei}
\begin{eqnarray}
v = \lim_{t \to +\infty} \frac{1}{t} \,
\int_{0}^{t}  \dot x(\tau) \, d\tau = \lim_{t \to +\infty} \frac{x(t)}{t},
\label{eqav}
\end{eqnarray}
independent of the initial conditions \cite{footnote}, 
is nonzero. Consider two temporal transformations: \emph{time reversal}
($\mathcal{R} \,:\,t\mapsto -t$) and \emph{time shift} ($\mathcal{S}\,:\,
t\mapsto t+T/2$), and suppose that their action on the force $f(t)$ is
given by
\begin{eqnarray}
&& (\mathcal{R}f)(t)= f(-t)=f(t),
\label{reversal} \\
&& (\mathcal{S}f)(t)=f(t+T/2)=-f(t).
\label{shift}
\end{eqnarray}
Suppose further that any of these transformations ---with the appropriate
transformation of $x(t)$--- leaves $E[x(t),f(t)]=0$ invariant. Non-dissipative
systems provide typical examples of this kind of behavior such as the equation of 
motions of cold atoms in optical lattices \cite{schia}, the dynamic of a particle in a symmetric potential 
 \cite{asym2}, and the soliton ratchets in the extended systems \cite{sazo,morales}.

For these kind of systems and forces which satisfy either (\ref{reversal}) 
or (\ref{shift}) ---or both--- there can be no ratchet effect because any of
the two transformations changes the sign of $v$ ($\mathcal{R}$ because time goes
backwards, and $\mathcal{S}$, because $f$ changes sign). As a matter of fact,
this is a nice illustration of Curie's principle \cite{curie}. 

In some other cases, time reversal leaves the equation invariant provided
the force transforms itself as
\begin{equation}
(\mathcal{R}f)(t)= f(-t)=-f(t)
\label{antireversal}
\end{equation}
instead of (\ref{reversal}). The most prominent examples of this are equations
describing overdamped systems such as the vortex motion in JJs \cite{nori}, the overdamped brownian 
motion \cite{rei2}, and 
the ratchet dynamics of breathers in the discrete Schr\"odinger equation \cite{gorbach}. 

Again, and for the same
reason, no ratchet effect can appear if the driving force fulfills
(\ref{antireversal}). In this case, however, in general breaking (\ref{antireversal}) and
(\ref{shift}) is not enough to induce a ratchet current, some additive noise
is necessary as well \cite{marchesoni86}.

Whichever the case, a bi-harmonic force like \eqref{eqf} is able to
break both (\ref{reversal}) [or (\ref{antireversal})] and (\ref{shift})
and induce a ratchet current. In what follows
we will prove that, provided a ratchet current is produced, the symmetries
impose that it be of the form \eqref{eq:vexp}.

Let us begin by noticing that $v$ must be a functional of $f(t)$, which we
can expand as
\begin{equation}
\begin{split}
v[f] &=v_0+\sum_{n=1}^{\infty}v_n[f], \\
v_n[f] &=\langle c_n(t_1,\dots,t_n)f(t_1)\cdots f(t_n)\rangle,
\end{split}
\label{eq:expansion}
\end{equation}
where
$\langle X\rangle\equiv T^{-n}\int_0^Tdt_1\cdots
\int_0^T dt_n\,X$ and
$v_0=v[0]$.
This functional Taylor expansion is a rigorous result of functional
analysis valid for a very wide class of functionals on Banach spaces
(see \cite{lusternik,binney,hansen} for details).
As $v[-f]=-v[f]$ for any force $f(t)$, $c_{2n}(t_1,\dots,t_{2n})\equiv 0$,
so only odd terms appear in the expansion (\ref{eq:expansion}). On the
other hand, the functions $c_n(t_1,\dots,t_n)$ can be taken 
$T$-periodic in each variable, and can always be chosen totally symmetric
under any exchange of their arguments. Notice in passing that only if 
$c_n(t_1,\dots,t_n)\propto\delta(t_{1}-t_{2})\cdots\delta(t_{n-1}-t_{n})$
can $v$ be expanded in moments of $f(t)$ ---thus the moment expansion is
only a particular case of \eqref{eq:expansion}.

Let us now specialize (\ref{eq:expansion}) for the bi-harmonic force
(\ref{eqf}). First of all, $v$ is not affected by the choice of
time origin; thus $v[\mathcal{T}_{\tau}f]=v[f]$, where $(\mathcal{T}_{\tau}f)(t)=f(t+\tau)$ 
for any $\tau$. 
But $f(t+\tau)=\epsilon_1\cos(qt+\tilde\phi_1)+\epsilon_2\cos(pt+\tilde\phi_2)$, with 
$\tilde\phi_1=\phi_1+q\tau$ and $\tilde\phi_2=\phi_2+p\tau$, so $v[f]$
must depend on the phases only through the combination 
$\theta=p\phi_1-q\phi_2=p\tilde\phi_1-q\tilde\phi_2$. 

Now we must compute $v_n[f]$ for any odd $n>0$. By expanding (\ref{eqf}) in
complex exponentials,
\begin{equation}
\begin{split}
v_n[f]=&\,\sum_{|{\bf n}|=n,\ {\bf n}\ge{\bf 0}} A({\bf n})
\epsilon_1^{n_1+n_2}\epsilon_2^{n_3+n_4} \\
&\times e^{i[(n_1-n_2)\phi_1+(n_3-n_4)\phi_2]},
\end{split}
\label{eq:sum}
\end{equation}
where ${\bf n}=(n_1,n_2,n_3,n_4)$, $|{\bf n}|\equiv n_1+n_2+n_3+n_4$, and
${\bf n}\ge{\bf 0}$ denotes a componentwise inequality. Besides,
because of the symmetry of the functions $c_n(t_1,\dots,t_n)$,
\begin{equation}
A({\bf n}) = \frac{n!2^{-n}}{\prod_{i=1}^4n_i!}
\left\langle c_n(t_1,\dots,t_n) e^{i \omega {\bf v}\cdot(t_1,\dots,t_n)} \right\rangle,
\label{eq:An}
\end{equation}
where
\begin{equation*}
{\bf v}\equiv(\overbrace{q,\dots,q}^{n_1},\overbrace{-q,\dots,-q}^{n_2},
\overbrace{p,\dots,p}^{n_3},\overbrace{-p,\dots,-p}^{n_4}).
\end{equation*}
The complex number $A({\bf n})=B({\bf n})e^{i\psi({\bf n})}$, where
\begin{equation}
\begin{split}
\psi(n_1,n_2,n_3,n_4) &= -\psi(n_2,n_1,n_4,n_3), \\
B(n_1,n_2,n_3,n_4) &= B(n_2,n_1,n_4,n_3).
\end{split}
\label{eq:psiB}
\end{equation}

Let us focus now on the complex exponential in (\ref{eq:sum}). We know that
$v$ must be a function of $\theta$, so the only nonzero terms in this sum
are those satisfying $(n_1-n_2)\phi_1+(n_3-n_4)\phi_2=k\theta$ for some
$k\in\ZZ$, i.e.\ $n_1-n_2=kp$ and $n_4-n_3=kq$.
But that means $|k|(p+q) =|n_1-n_2-n_3+n_4|
\le |n_1+n_2+n_3+n_4|=n$.

Suppose $n<p+q$; then $k=0$, which implies $n_1=n_2$ and $n_3=n_4$, and
therefore $n$ must be even. Thus $v_n[f]$ must be zero for any odd $n<p+q$,
but since there are no even terms in (\ref{eq:expansion}), this means 
that no term with $n<p+q$ contributes to $v$.

Suppose now $n\ge p+q$; then $|k|>0$ ($k=0$ is excluded because it would lead
to an even $n$) and so there will be nonzero terms in (\ref{eq:sum}) corresponding
to powers $\epsilon_1^{2n_2+|k|p}\epsilon_2^{2n_3+|k|q}$ or
$\epsilon_1^{2n_1+|k|p}\epsilon_2^{2n_4+|k|q}$. The lowest order is $n=p+q$,
and is obtained either when $n_2=n_3=0$, $n_1=p$, and $n_4=q$ (i.e.\ $k=1$),
or when $n_1=n_4=0$, $n_2=p$, and $n_3=q$ (i.e.\ $k=-1$).
Because of (\ref{eq:psiB}), the contribution of these two terms to (\ref{eq:sum}) is
\begin{equation}
v_{p+q} = B\epsilon_1^p\epsilon_2^q\cos(\theta+\theta_0),
\end{equation}
where $B=2B(p,0,0,q)$ and $\theta_0=\psi(p,0,0,q)$. 

Let us now assume that the equation is invariant when the force satisfies
(\ref{reversal}).
Then, since $\langle c_n(t_1,\dots,t_n)f(-t_1)\cdots f(-t_n)\rangle=
\langle c_n(-t_1,\dots,-t_n)f(t_1)\cdots f(t_n)\rangle$, and $v[\mathcal{R}f]
=-v[f]$, hence 
\begin{equation}
c_n(-t_1,\dots,-t_n)=-c_n(t_1,\dots,t_n).
\end{equation}
Applied to (\ref{eq:An}) this means that $B({\bf n})e^{-i\psi({\bf n})}
=-B({\bf n})e^{i\psi({\bf n})}$, i.e.\ $\psi({\bf n})=\pi/2$ for all ${\bf n}$.

On the other hand, if the equation is invariant when the force satisfies
(\ref{antireversal}), then $v[-\mathcal{R}f]=-v[f]$, so (recall that $n$ is
odd)
\begin{equation}
c_n(-t_1,\dots,-t_n)=c_n(t_1,\dots,t_n).
\end{equation}
Applied to (\ref{eq:An}) this means that $B({\bf n})e^{-i\psi({\bf n})}
=B({\bf n})e^{i\psi({\bf n})}$, i.e.\ $\psi({\bf n})=0$ for all ${\bf n}$.

What we have just shown is that the mean velocity, $v$, of the ratchet current
induced by the bi-harmonic force (\ref{eqf}) in an equation which is
invariant under (\ref{shift}) always has the form \eqref{eq:vexp}
if the amplitudes $\epsilon_1$ and $\epsilon_2$ are small.  
The coefficients $B$ and $\theta_{0}$ 
depend on the frequency and on the remaining parameters
of the system, but not in a universal way that can be predicted under
symmetry arguments like these ones.
This shape for the current has been observed, mostly for $p=2$ and $q=1$,
in experimental, numerical, and theoretical results 
in several seemingly unrelated systems 
\cite{usti,schia,asym2,marchesoni86,rei3,sazo,morales,new27b}.  
For $p=4$ and $q=1$, the collective coordinate
on soliton ratchets developed in \cite{chaos} also confirms
\eqref{eq:vexp}.

If the equation
is also invariant under (\ref{reversal}) \cite{schia,asym2}, 
then $\theta_0=\pi/2$ and we
recover the form
$v\sim\epsilon_1^p\epsilon_2^q\sin\theta$,
whereas if the equation is invariant under
(\ref{antireversal}), then $\theta_0=0$ and
$v\sim\epsilon_1^p\epsilon_2^q\cos\theta$,
in agreement with the vortex motion observed in JJs \cite{nori},
with the overdamped stochastic dynamic of particles studied in
\cite{brownian,marchesoni86}, and with
the ratchet mobility of breathers in the discrete nonlinear 
Schr\"odinger equation computed for $p=2$ and $q=1$ in \cite{gorbach}.      

Notice that formula (\ref{eq:vexp}) does not imply that $B$ must 
be nonzero (Curie's principle). It only proves, under symmetry arguments,
that the leading term of $v$ can be of that precise form. It might well happen,
for some specific equation, that $B=0$. In this case this analysis
shows that the leading term must have a dependence on the amplitudes through
powers higher than $p$ and $q$. It is likely that if this occurs it will be
the fingerprint of a hidden symmetry which, properly broken, will restore
the result (\ref{eq:vexp}). 

This analysis 
provides a direct 
way to quantitatively relate the causes  
and the consequences of phenomena through Curie's principle. 
For instance, our study can be extended to the so-called 
gating effect, i.e.\ when  the amplitude of spatial or 
field potentials for particles or solitons, respectively,
is modified by a multiplicative force $g(t)=\epsilon_{2} \cos(p \omega t
+ \phi_2)$ as well as an additive force, $f(t)=\epsilon_{1} \cos(q \omega t +\phi_1)$,
with $p$ and $q$ coprimes, acts on the system \cite{gating,gating2,chaos-marche}.
In such systems, if both $f(t)$ and $g(t)$ satisfy (\ref{reversal}) 
(in the non-damped limit) or (\ref{antireversal}) (in the overdamped limit),
or $f(t)$ fulfills (\ref{shift}),  
a ratchet transport cannot be
induced. A similar procedure shows that, when these symmetries are broken,
the average ratchet velocity is also given by Eq.~(\ref{eq:vexp}), where
$\theta_0=0$ or $\theta_0=\pi/2$ in the non-damped or overdamped 
limits, respectively \cite{nos}.      

We acknowledge financial support through grants MTM2006-13000-C03-01 (RAN),
FIS2008-02380/FIS (NRQ), and MOSAICO (JAC) (from Ministerio de Educaci\'on y
Ciencia, Spain), grants FQM262 (RAN), FQM207 (NRQ), and P06-FQM-01735 (NRQ, RAN)
(from Junta de Andaluc\'{\i}a, Spain), and project MOSSNOHO-CM (JAC) (from
Comunidad de Madrid, Spain).


\begin{thebibliography}{88}

\bibitem{ieee} Yu.\ K.\ Pozhela and H.\ J.\ Karlin, Proc.\ IEEE \textbf{53}, 1788 (1965). 

\bibitem{new28} W.\ Schneider and K.\ Seeger, Appl.\ Phys.\ Lett.\ \textbf{8}, 133 (1966). 

\bibitem{epl99} K.\ N.\ Alekseev, M.\ V.\ Erementchouk and F.\ V.\ Kusmartsev, Europhys.\ 
Lett.\   \textbf{47}, 595 (1999). 

\bibitem{usti}
A.\ V.\ Ustinov, C.\ Coqui,
A.\ Kemp, Y.\ Zolotaryuk, and M.\ Salerno,
Phys.\ Rev.\ Lett.\ \textbf{93}, 087001 (2004).  

\bibitem{nori}
S.\ Ooi, S.\ Savel'ev, M.\ B.\ Gaifullin, T.\ 
Mochiku, K.\ Hirata, and F. Nori, Phys.\ Rev.\ Lett.\ \textbf{99}, 207003 (2007).

\bibitem{vicent}
J.\ E.\ Villegas, S.\ Savel'ev, F.\ Nori, E.\ M.\ Gonz\'alez,
J.\ V.\ Anguita, R.\ Garc\'{\i}a, J.\ L.\ Vicent, Science \textbf{302},
1188 (2003).

\bibitem{schia}
M.\ Schiavoni, L.\ S\'{a}nchez-Palencia, F.\ Renzoni, and
G.\ Grynberg, Phys.\ Rev.\ Lett.\ \textbf{90}, 094101 (2003).

\bibitem{gommers}
R.\ Gommers, S.\ Bergamini, and F.\ Renzoni, 
Phys.\ Rev.\ Lett.\ \textbf{95}, 073003 (2005).

\bibitem{mm}
M.\ O.\ Magnasco, Phys. Rev. Lett. \textbf{71},
1477 (1993). 

\bibitem{app}
P.\ Reimann and P.\ H\"anggi, Appl.\ Phys.\ A
\textbf{75}, 169 (2002).

\bibitem{brownian}
P.\ H\"anggi, F.\ Marchesoni, and F.\ Nori,
Ann.\ Phys.\ (Leipzig) \textbf{14}, 51 (2005).

\bibitem{physicstoday}
R.\ D.\ Astumian and P.\ H\"anggi, Phys.\ Today  \textbf{55}, 32 (2002).

\bibitem{rei} 
P.\ Reimann, Phys.\ Rep.\ \textbf{361}, 57 (2002).

\bibitem{adjari}
A.\ Ajdari, D. Mukamel, L.\ Peliti, and J.\ Prost,
J.\ Phys.\ I France \textbf{4}, 1551 (1994).

\bibitem{asym2} S.\ Flach, O.\ Yevtushenko, and Y.\ Zolotaryuk,
Phys.\ Rev.\ Lett. \textbf{84}, 2358 (2000).

\bibitem{rei2}
P.\ Reimann, Phys.\ Rev.\ Lett.\ \textbf{86}, 4992 (2001).

\bibitem{sazo} 
M.\ Salerno and Y.\ Zolotaryuk, 
Phys.\ Rev.\ E \textbf{65}, 056603 (2002).

\bibitem{cole}
D.\ Cole, S. Bending, S.\ Savel'ev, A.\ Grigorenko, T.\ Tamegai, and
F.\ Nori, Nat.\ Mater.\ \textbf{5}, 305 (2006).

\bibitem{renzoni3}
R.\ Gommers, P.\ Douglas, S.\ Bergamini, M.\ Goonasekera,
P.\ H.\ Jones, and F.\ Renzoni, Phys.\ Rev.\ Lett.\
\textbf{94}, 143001 (2005).  

\bibitem{taylor} C.\ E.\ Skov and E.\ Pearlstein, Rev.\ Sci.\ Inst.\
\textbf{35}, 962 (1964).  

\bibitem{marchesoni86}
F.\ Marchesoni, Phys.\ Lett.\ A \textbf{119}, 221 (1986).

\bibitem{new27b} O.\ Yevtushenko, S.\ Flach, Y.\ Zolotaryuk, and A.\ A.\ Ovchinnikov,  
Europhys.\ Lett.\ \textbf{54}, 141 (2001).

\bibitem{rei3}
A.\ Engel, H.\ W.\ M\"uller, P.\ Reimann, 
A.\ Jung, Phys.\ Rev.\ Lett.\ \textbf{91}, 060602 (2003). 

\bibitem{goychuk}
I.\ Goychuk, P.\ H\"anggi, Europhys.\ Lett.\ \textbf{43}, 503 (1998).

\bibitem{morales}
L.\ Morales-Molina, N.\ R.\ Quintero, F.\ G.\ Mertens, and A.\ S\'anchez,
Phys.\ Rev.\ Lett. \textbf{91}, 234102 (2003).

\bibitem{chaos}
L.\ Morales-Molina, Niurka R.\ Quintero,
A.\ S\'anchez, and F.\ G.\ Mertens, Chaos \textbf{16}, 013117 (2006).

\bibitem{gorbach}
A.\ V.\ Gorbach, S.\ Denisov, S.\ Flach, Optics Lett.\  
\textbf{31}, 1702 (2006).

\bibitem{taylor2}
S.\ Denisov, S.\ Flach, A.\ A.\ Ovchinnikov, O.\ Yevtushenko, and Y.\ 
Zolotaryuk, Phys.\ Rev.\ E \textbf{66} 041104 (2002).

\bibitem{footnote} The definition of $v$ is independent on the initial conditions 
for any dissipative system.
Nondissipative systems are idealizations impossible to realize experimentally. They can only be
approached as the limit of very weak damping \cite{gommers,new27b}. In this sense,
definition (\ref{eqav}) remains valid in this limit.


\bibitem{curie}
P.\ Curie, J.\ Phys.\ (Paris) S\'er.\ 3, {\bf III}, 393 (1894).

\bibitem{lusternik}
R.\ F.\ Curtain and A.\ J.\ Pritchard, \emph{Functional Analysis in Modern
Applied Mathematics} (Academic Press, London, 1977), theorem~6.4, p.~101.

\bibitem{binney}
J.\ J.\ Binney, N.\ J.\ Dowrick, A.\ J.\ Fisher, and M.\ E.\ J.\ Newman,
\emph{The Theory of Critical Phenomena: An Introduction to the Renormalization Group}
(Oxford University Press, Oxford, 1992), Appendix L, p.~404.

\bibitem{hansen}
J.\ P.\ Hansen and I.\ R.\ McDonald, \emph{Theory of Simple Liquids}, 3rd ed.\
(Academic Press, London, 2006), formula 3.2.24, p.~53.

\bibitem{chaos-marche}
M.\ Borromeo and F.\ Marchesoni, Chaos \textbf{15}, 026110 (2005).

\bibitem{gating2} 
E.\ Zamora-Sillero, N.\ R.\ Quintero, and F.\ G.\ Mertens, 
Phys.\ Rev.\ E \textbf{74} 046607  (2006).

\bibitem{gating} R.\ Gommers, V.\ Lebedev, M.\ Brown, and F. Renzoni,  
Phys.\ Rev.\ Lett.\ \textbf{100}, 040603  (2008).

\bibitem{nos} N.\ R.\ Quintero,  R.\ Alvarez-Nodarse, and J.\ A.\ Cuesta,
in preparation.



\end{thebibliography}
\end{document}


\title{Symmetries shape the current in ratchets induced by a bi-harmonic force.
Supplementary Material}

\author{Niurka R.\ Quintero}

\author{Jos\'e A.\ Cuesta}

\author{Renato Alvarez-Nodarse}

\maketitle

Let us analyze the following evolution equations $E[x(t),f(t)]=0$ for the
variables $x(t)$ (position) and $u(t)$ (velocity) of a relativistic particle
of mass $M>0$  
\bq\begin{split} \label{equx}
M \frac{du}{dt} & = - f(t) (1-u^2)^{3/2}-\gamma u(1-u^2),\\
\frac{dx}{dt} & = u(t), \qquad u(0)=u_{0}, \quad x(0)=x_{0}, 
\end{split}
\eq
where $x_{0}$ and $u_{0}$ are the initial conditions, $\gamma>0$ represents 
the damping coefficient and $f(t)$ is a $T$-periodic driving force \cite{olsen:1983}.    
Notice that defining the momentum
\bq\label{def-P}
P(t)=\frac{M u(t)}{\sqrt{1-u^2(t)}}, 
\eq
we can transform Eq.~\eqref{equx} into the linear equation
\bq\label{eq-P}
\frac{dP}{dt}=-\beta P-f(t),
\eq
where $\beta=\gamma/M$, whose solution is given by
\begin{equation} \label{solu-P}
P(t) = P(0) e^{-\beta t} - \int_{0}^{t} \, dz f(z) e^{-\beta (t-z)}. 
\end{equation}
Equation~(\ref{equx}) is invariant under \emph{time shift} ($\mathcal{S}\,:\,
t\mapsto t+T/2$) along with the change $x \mapsto -x$, provided
$(\mathcal{S}f)(t)=f(t+T/2)=-f(t)$. The bi-harmonic force
\begin{eqnarray}
f(t)=\epsilon_{1} \cos(q \omega t +\phi_{1})+
\epsilon_{2} \cos(p \omega t +\phi_{2}), 
\label{eq-f}
\end{eqnarray}
preserves this symmetry if, both, $p$ and $q$ are odd integer numbers, so
in this case the average velocity 
\begin{eqnarray}
v = \lim_{t \to +\infty} \frac{1}{t} \,
\int_{0}^{t} u(\tau) \, d\tau, 
\label{eqav}
\end{eqnarray}
is zero. In contrast, if $p+q$ is odd and $p$ and $q$ are coprimes, a nonzero
average current can appear. For the sake of simplicity we will take
$p=2$ and $q=1$ in Eq.~\eqref{eq-f} \cite{sazo}. Then the solution to
\eqref{solu-P} for the chosen force \eqref{eq-f} will be
\begin{eqnarray}\nonumber
\ P(t)  &=& \tilde{P}_{0} \exp(-\beta t)-\frac{\epsilon_{1} }{\sqrt{\beta^2+\omega^2}}
\cos(\omega t+\phi_{1}-\chi_{1}) \\ \label{2.6pop}
 &-&\frac{\epsilon_{2} }{\sqrt{\beta^2+4 \omega^2}} \cos(2 \omega t+\phi_{2}-\chi_{2}),
\end{eqnarray}
with   $\tilde{P}_{0}=P(0)+(\epsilon_{1}/\sqrt{\beta^2+\omega^2}) \cos(\phi_{1}-\chi_{1}) + 
(\epsilon_{2}/\sqrt{\beta^2+4 \omega^2})  \cos(\phi_{2}-\chi_{2})$, 
$\chi_{1}=\arctan\left(\omega/\beta\right)$, and  
$\chi_{2}=\arctan\left(2 \omega/\beta\right)$. 
From (\ref{def-P}), one obtains
\begin{eqnarray} 
\label{eq-u1}
u(t) &= & \sum_{k=0}^{\infty} \frac{(-1)^{k} (1/2)_{k}}{k! M^{2k+1}} 
[P(t)]^{2 k +1},  
\end{eqnarray}
where $(1/2)_{k}\equiv (1/2)(1/2+1)\cdots(1/2+k-1)$.
From \eqref{eqav} and \eqref{eq-u1} it follows that the time-average velocity, $v$,  
cannot be expressed as a function of the odd moments of $f(t)$, unless
$P(t)$ is proportional to $f(t)$. Indeed, it is only in the overdamped
case [in which the inertial term in \eqref{equx} is neglected] that the
evolution equation is given by $P(t)=-(1/\beta) f(t)$ and then $v$ do
admit an expansion in odd moments of $f(t)$.

Moreover, for small amplitudes $\epsilon_{1}$ and $\epsilon_{2}$, 
the leading term of the time-average velocity (\ref{eq-u1}) reads 
\bq\label{vmedia-1}\begin{split}
v = B \epsilon_1^2\epsilon_2  \cos(2\phi_1-\phi_2+\theta_{0}),   
\end{split}
\eq
where $B=3/(8 M^3 (\beta^2+\omega^2) \sqrt{\beta^2+4\omega^2})$
and $\theta_{0}=-2\chi_1+\chi_2$.  This expression is in agreement with
the prediction of our theory. Furthermore, in the limit $\beta \to 0$
we have $-2\chi_1+\chi_2\to\pi/2$, and in the combined limit $M \to 0$
and $\beta \to \infty$, with $\gamma=\text{const.}$, $-2\chi_1+\chi_2\to 0$.
One can check that in the former case Eq.~\eqref{equx} is invariant
under \emph{time reversal} ($\mathcal{R} \,:\,t\mapsto -t$) provided
$(\mathcal{R}f)(t)= f(-t)=f(t)$, and therefore $\theta_{0}=\pi/2$ is
the prediction of our theory. In the latter case, however, it is
$(\mathcal{R}f)(t)=f(-t)=-f(t)$ that leaves Eq.~\eqref{equx} invariant
and then our theory predicts $\theta_{0}=0$.